\documentstyle[preprint,aps]{revtex}
\begin{document}
\draft

\title{\bf
Hyperuniversality of Fully Anisotropic Three-Dimensional
Ising Model}
\author{M. A. Yurishchev}
\address{
Radiophysical Research Institute, 603600 Nizhny Novgorod, Russia}

\date{\today}
\maketitle

\begin{abstract}
For the fully anisotropic simple-cubic Ising lattice, the critical
finite-size scaling amplitudes of both the spin-spin and
energy-energy inverse correlation lengths and the singular part
of the reduced free-energy density are calculated by the
transfer-matrix method and a finite-size scaling for cyclic
$L\times L\times\infty$ clusters with $L=3$ and 4.
Analysis of the data obtained shows that the ratios and the
directional geometric means of above amplitudes are universal.
\end{abstract}



\section{Introduction}
Conforming with the Privman-Fisher hyperuniversality hypothesis,
the finite-size scaling (FSS) equations for the inverse correlation
lengths and the singular part of the reduced free-energy density
near the bulk phase transition of a system have, respectively,
the form (for reviews see Refs.~\cite{1,2})
\begin{equation}
   \kappa_{i,L}(t, h)
   = L^{-1}X_i(C_1tL^{y_T}, C_2hL^{y_h})
\end{equation}
and
\begin{equation}
   f_L^{(s)}(t, h) = L^{-d}Y(C_1tL^{y_T}, C_2hL^{y_h})\;.
\end{equation}
Here $L$ is a characteristic size of finite or partly finite
subsystem, the index $i$ labels the types of correlation lengths
[spin-spin ($i=1$), energy-energy ($i=2$), etc], $d$ is the space
dimensionality, $t=(T-T_c)/T_c$, $h$ is an external field, $y_T$
and $y_h$ are the critical exponents, $X_i(x,y)$ and $Y(x,y)$
are the scaling functions which, within the limits of universality
classes, can else depend on the type of boundary conditions and
the subsystem shape; all non-universality of a model is absorbed
in the metric factors $C_1$ and $C_2$.
Equations (1) and (2) allow to find the universal combinations
for the FSS amplitudes at the phase-transition point $t=h=0$.
In particular, the amplitudes for the inverse correlation
lengths $A_s=X_1(0,0)$ and $A_e=X_2(0,0)$ and for the free energy
$A_f=Y(0,0)$ must be universal
themselves.
In the case of strips with periodic boundary conditions,
they are (see, e.\ g., Ref.~\cite{2})
\begin{equation}
   A_s=\pi\eta,\qquad A_e=\pi\eta_e,\qquad A_f=\frac{\pi c}{6},
\end{equation}
where $\eta$ and $\eta_e$ are the exponents of the decay law
correspondingly of the spin-spin and energy-energy correlation
functions ($\eta=1/4$ and $\eta_e=2$ for the flat Ising model) and
$c$ is the central charge of Virasoro algebra ($c=1/2$ for the
two-dimensional Ising lattice).

All foregoing statements are applied to the {\it spatially
isotropic\/} systems.
Lattice anisotropy is a marginal effect and hence the amplitudes
and their combinations, strictly speaking, must depend on
anisotropy parameters~\cite{2}.
However in the case of the anisotropic two-dimensional Ising model,
it has been established~\cite{3,4} that although the inverse
correlation-lengths and free-energy amplitudes get a
non-universal factor, $R_{\alpha}$ ($\alpha$ labels the directions
along which an $L\times\infty$ strip is infinite; here,
$\alpha=x, z$), it is common and the directional geometric mean
$\bar R=(R_xR_z)^{1/2}$ is a constant (equalling the unity).
Therefore, the universality is preserved for the ratios and
the directional geometric means of these amplitudes.

In the light of above, it would be interesting to clear the
matter up in three dimensions.
Such attempt is undertaken in the present paper.
We consider the three-dimensional Ising model on a simple-cubic
lattice with different interaction constants $J_x$, $J_y$, and
$J_z$ along all three spatial directions.
The lattice is approximated by the $L\times L\times\infty$
bars with periodic boundary conditions in both transverse
directions.
Such boundaries eliminate undesirable surface effects and hence
improve a quality of approximation.
By the transfer matrix (TM) method combined with FSS analysis
for the subsystems with sizes $L=3$ and 4, we determine at first
the critical temperatures depending upon anisotropy parameters
$J_x/J_z$ and $J_y/J_x$.
(We consider a system at least with two non-zero couplings;
unless otherwise stated, the $L\times L\times\infty$ parallelepipeds
are taken infinitely long in the $z$ direction.)
After this, the FSS amplitudes of the inverse correlation lengths
and the free energy are calculated at the critical points found.
The obtained results demonstrate the independence of the amplitude
ratios on the parameter $J_x/J_z$ when $J_y/J_x$ is fixed.
Moreover, the analysis shows that the ratios are also independent
on the second anisotropy parameter $J_y/J_x$ at any rate in the
region $J_y/J_x\simeq 1$.
Finally, our calculations give evidence in the constancy of the
directional geometric mean of the spin-spin inverse correlation
length amplitude in three dimensions.
Together with an invariance of the ratios, this implies that the
directional geometric means of other amplitudes must be
universal also.


\section{Model and Solution of the Eigenproblems}
The Hamiltonian of Ising model on a simple-cubic lattice with
nearest-neighbor interactions reads
\begin{equation}
   {\cal H}=-\sum_{ijk}S_{ijk}(J_xS_{i+1jk}
   +J_yS_{ij+1k}+J_zS_{ijk+1})\;.
\end{equation}
The spin-field variables  $S_{ijk}$ are located in the lattice
sites and take the values $\pm 1$.

The transfer matrix $V$ of an $L\times L\times\infty$ subsystem
is introduced by elements
\begin{eqnarray}
   \langle S_{11},S_{12},\ldots,S_{LL}|V|& &
   S_{11}^\prime,S_{12}^\prime,\ldots,S_{LL}^\prime
   \rangle=\prod_{i,j=1}^L
   \exp[\case{1}{2}K_x(S_{ij}S_{i+1j}
   +S_{ij}^\prime S_{i+1j}^\prime)\nonumber\\
   & &
   +\case{1}{2}K_y(S_{ij}S_{ij+1}+
   S_{ij}^\prime S_{ij+1}^\prime)
   +K_zS_{ij}S_{ij}^\prime]\;,
\end{eqnarray}
where $K_\alpha = J_\alpha/k_BT$ (now $\alpha = x, y, z$);
$S_{iL+1}=S_{i1}$ and $S_{L+1j}=S_{1j}$ by all
$i,j=1,2,\ldots,L$.
The matrix $V$ is real, symmetric and has an order of $2^N$
where $N=L^2$ equals the number of chains in a system;
that is dense and all its elements are positive.

The principal task is to find the eigenvalues of $V$ because,
for example, the density of a free energy measured in units
of $-k_BT$ is given by
\begin{equation}
   f_L=N^{-1}\ln \Lambda_0\;,
\end{equation}
where $\Lambda_0$ is the largest eigenvalue of a TM.
The inverse longitudinal correlation lengths (mass gaps) equal
\begin{equation}
   \kappa_{i,L}=\ln(\Lambda_0/\Lambda_i)\;,
\end{equation}
where $\Lambda_1, \Lambda_2\ldots$ are the next (after
$\Lambda_0$) dominant eigenvalues of TM for the subsystem.

In order to solve the TM eigenproblem for $L$ as large as
possible, we reduce the TMs to the block-diagonal forms
using a symmetry under the transformations
of the group ${\rm Z}_2\times{\rm T}\wedge{\rm C}_{2\rm v}$.
Here ${\rm Z}_2$ is a group of global spin inversions
$S\rightarrow -S$, ${\rm T}$ is a group of translations in
the transverse directions of a bar, and ${\rm C}_{2\rm v}$ is
the point group consisting of rotations around the axis of a
subsystem at angles multiple to $\pi$ and the
reflections in planes going through this axis and the
middles of opposite sides of an $L\times L\times\infty$
parallelepiped.

There is no necessity to perform the full quasidiagonalization
of TMs because the leading eigenvalues are distributed only
among two subblocks.
Owing to the Perron theorem~\cite{5}, $\Lambda_0$ lies in the
subblock of an identity irreducible representation.
$\Lambda_1$ is located in the other subblock --- it is built on
the basis functions which are symmetrical under all
transformations of the space subgroup
${\rm T}\wedge{\rm C}_{2\rm v}$ and antisymmetrical under the
transformations including a spin inversion.
$\Lambda_2$ is situated again in the subblock of an identity
irreducible representation.
(In connection with this see, for example, Ref.~\cite{3}.)

As a group-theoretical analysis shows (see Appendix A), both subblocks
containing the largest eigenvalues have sizes of $18\times18$
in the case of $3\times3\times\infty$ cluster.
For a cylinder $4\times4\times\infty$, the TM
$65\,536$ by $65\,536$ is reduced to a block-diagonal form in
which the required subblocks have the orders 787 and 672.
The final extraction of needed eigenvalues of TMs was carried
out by a numerical solution of eigenproblems for the corresponding
subblocks.
By this, we applied the conjugate gradient method~\cite{6}
and, if necessary, used also the library functions {\it tred2\/}
and {\it tqli\/}~\cite{7}.
Calculations were run on IBM PC-486 computer in the operating
system LINUX.


\section{Calculation of the Critical Amplitudes}
So, the FSS amplitudes for the inverse correlation lengths of the
spin-spin and energy-energy correlation functions are equal to
\begin{equation}
   A_s=L\kappa_{1,L}
\end{equation}
and
\begin{equation}
   A_e=L\kappa_{2,L}\;,
\end{equation}
where $\kappa_{1,L}$ and $\kappa_{2,L}$ have been taken at the
phase-transition point $T_c$ (by $h=0$).
This point itself was determined from the renormalization-group
equation
\begin{equation}
   L\kappa_{1,L}(T_c)=
   (L-1)\kappa_{1,L-1}(T_c)
\end{equation}
with $L=4$.
The amplitude for the singular part of a free-energy density,
$A_f$, is found from a system of equations
\begin{equation}
   f_L=f_0+L^{-d}A_f
\end{equation}
with $L=3$ and 4.
Here $f_0$ denotes the regular (background) part of a
free-energy density; $f_3$ and $f_4$ are taken again at the
critical points.

The critical temperatures, amplitudes, and background $f_0$
calculated at different values of the anisotropy parameters
$J_x/J_z$ and $J_y/J_x$ are collected in table I.
In Eqs.~(11), the spatial dimensionality has been put $d=2$
for $J_y=0$ and $d=3$ for $J_y\neq0$.
It should be noted also that, for finite $L$, Eq.~(7) leads to the
wrong values for $\kappa_{2,L}$ in the limit of non-interacting
strips ($J_y=0$).
Due to $\lambda_1^2>\lambda_0\lambda_2$
($\lambda_0>\lambda_1>\lambda_2$ are the largest eigenvalues
of a transfer matrix for the strip),
$\Lambda_2=(\lambda_0\lambda_1)^2$ by $L=4$ and therefore
$\kappa_{2,L}=\ln[\lambda_0^4/(\lambda_0\lambda_1)^2]=2\kappa_{1,L}$.
However, the correct values are given by formula
$\kappa_{2,L}=\ln(\lambda_0/\lambda_2)$ which has been used to
build up the table I.

In table II, we present the data for the directional geometric
mean of the spin-spin inverse correlation length amplitude
$\bar A_s$.
Calculations were performed by the equation
\begin{equation}
   \bar A_s = \left\{
   \begin{array}{ll}
   (A_s^{(x)}A_s^{(z)})^{1/2}            & if\ J_y = 0\\
   (A_s^{(x)}A_s^{(y)}A_s^{(z)})^{1/3}   & if\ J_y \neq 0
   \end{array}
   \right.,
\end{equation}
where $A_s^{(\alpha)}$ is the amplitude of the spin-spin
inverse correlation length when the bar $L\times L\times\infty$
was stretched (for given $J_x, J_y,$ and $J_z$) along
the $\alpha$ direction.


\section{Discussion}
Consider first the behavior of absolute amplitudes.
For the three-dimensional systems, available information
about them is very scanty.
In the periodic cylinder geometry, it seems to be known only the
estimates for the correlation-length amplitudes
found by Monte Carlo simulations on the fully isotropic
($J_x=J_y=J_z$) lattices $L\times L\times 128 $ with $L=4,6,8,$
and 10 (Ref.~\cite{8}).
For the inverse correlation-length amplitudes, these estimates
($L=10$) yield $A_s=1.342$ and $A_e=4.78$.
Appealing to table I, one can convince oneself that our
calculations conform with these values.
Note also that the available high-temperature series for the
free energy of a fully isotropic simple-cubic Ising lattice
yields~\cite{9} $f_0=0.77711$ at criticality.
Our estimate for the background, 0.773, is in good agreement with
this magnitude.

In the two-dimensional case ($J_y=0$), there exists, vice versa,
complete information concerning the FSS amplitudes for the
inverse correlation lengths and the free energy in the
rectangular lattice with arbitrary anisotropy \cite{3,4}:
\begin{equation}
   A_s=\frac{\pi}{4}\left[\,\frac{
   \sinh(2J_x/k_BT_c)}{
   \sinh(2J_z/k_BT_c)}\,\right]^{1/2}\ ,
\end{equation}
\begin{equation}
   A_e=2\pi\left[\,\frac{
   \sinh(2J_x/k_BT_c)}{
   \sinh(2J_z/k_BT_c)}\,\right]^{1/2}\ ,
\end{equation}
and
\begin{equation}
   A_f=\frac{\pi}{12}\left[\,\frac{
   \sinh(2J_x/k_BT_c)}{
   \sinh(2J_z/k_BT_c)}\,\right]^{1/2}\ ,
\end{equation}
where the critical temperature $T_c$ satisfies to the equation
\begin{equation}
   \sinh\left(\frac{2J_x}{k_BT_c}\right)
   \sinh\left(\frac{2J_z}{k_BT_c}\right)=1\;.
\end{equation}
Our numerical results reproduce these rigorous dependencies
with acceptable accuracy.
For the isotropic square Ising lattice, the critical free
energy is (see Ref.~\cite{10})
\begin{equation}
   f_0=2G/\pi+\case{1}{2}\ln 2=0.929\,695\ldots
\end{equation}
($G=1^{-2}-3^{-2}+5^{-2}-\ldots$ is Catalan's constant).
Appropriate value from table I ($f_0$ at $J_x=J_z$ and $J_y=0$)
agrees to within $1.7\%$ with the given exact quantity.

Inspecting table I, we see the amplitudes vary in wide limits
reaching several orders.
The behavior is changed into a contrary one for their ratios.
First what draws attention is that the ratios $A_e/A_s$ and
$A_f/A_s$ stay practically unchanged with variation of
$J_x/J_z$ on three orders ($1-10^{-3}$) by given $J_y/J_x$.
In the two-dimensional space ($J_y=0$), the mean (here and below, over
$J_x/J_z$) value of $A_f/A_s$ equalling to 0.331 conforms
with the true value 1/3; the mean of $A_e/A_s$
equals to 7.2 that agrees, in order of magnitude, with the
exact value, 8, for the $A_e/A_s$ [see Eqs.~(13) -- (15)].
For the three-dimensional lattice with $J_x=J_y$, the mean
value of $A_f/A_s$ is 0.288.
This quantity agrees with estimate $A_f/A_s=0.272$ which follows
from the calculations of relative amplitudes for the inverse
correlation lengths and the free energy in the Hamiltonian limit
of a three-dimensional Ising model (square lattices $L\times L$
with sizes $L$ up to 5)~\cite{11}.
According to table I, the ratio for the inverse correlation-length
amplitudes is $A_e/A_s=3.53(6)$ in the discussed case.
This estimate is in agreement with the mean values
$A_e/A_s=3.62(7)$, Ref.~\cite{12}, and $A_e/A_s=3.7(1)$,
Ref.~\cite{8}.
Thus, the amplitude ratios $A_e/A_s$ and $A_f/A_s$ are not only
universal with respect to the $J_x/J_z$ but also their values
agree quantitatively with available estimates in two limited
cases: $J_y/J_x=0$ and 1.

We now discuss the dependence on $J_y/J_x$ in the intermediate
region.
In the limit $J_y/J_x\rightarrow 0 $, the $L\times L\times\infty$
bar decomposes into $L$ of independent strips $L\times\infty$
and consequently the TM of the bar is factorized into the direct
product of TMs for the strips.
Since the TM of the bar is finite by finite $L$, its eigenvalues
are {\em continuous\/} functions of model parameters.
Hence there must exist the $d=3\rightarrow d=2$ transition
region when $J_y/J_x\rightarrow 0 $.
To estimate its sizes by using $L$, we have calculated the
critical exponents $\nu$ and $\gamma/\nu$.
The calculation was performed via the ordinary FSS formulae
(see, e.~g., Ref.~\cite{13}):
\begin{equation}
   \nu=\frac{\ln[L/(L-1)]}
   {\ln[L\kappa_{1,L}^\prime/(L-1)\kappa_{1,L-1}^\prime]}
\end{equation}
and
\begin{equation}
   \gamma/\nu=\frac{\ln(\chi_L/\chi_{L-1})}
   {\ln[L/(L-1)]}\;,
\end{equation}
in which we put $L=4$.
Here $\kappa_{1,L}^\prime$ is the derivative of $\kappa_{1,L}$
with respect to the temperature and $\chi_{L-1}$ and $\chi_L$
are the magnetic susceptibilities of subsystems at the
phase-transition point.
(Formulae for the susceptibilities are derived in Appendix B.)
How the calculation gives the critical exponents $\nu$ and,
especially, $\gamma/\nu$ are practically constants with respect
to $J_x/J_z$ ($=1-10^{-3}$).
Their dependences on $J_y/J_x$ are shown in Fig.~1.
Within the section $0.2<J_y/J_x\leq 1 $, the exponents $\nu$ and
$\gamma/\nu$ preserve the unchanged values equalling,
respectively, to 0.67 and 1.97 that agrees with available estimates
for these exponents in the case of the fully isotropic
three-dimensional Ising model (Ref.~\cite{14} and references
therein).
By $J_y=0$, our calculation yields $\nu=1.06$ and
$\gamma/\nu=1.74$.
These magnitudes conform closely with the exact values of
discussed exponents in two dimensions: $\nu=1$ and
$\gamma/\nu=7/4$.
In Fig.~1, it is clear-cut displayed the region
$0\leq J_y/J_x<0.1 - 0.2$ where a smooth transition occurs from
the $d=3$ exponent values to the $d=2$ ones.
Consequently, one does not consider the $L\times L\times\infty$
lattice with $L\leq4$ as a three-dimensional one when
$J_y/J_x<0.2$.
In order to support this conclusion, we have calculated
the ``effective'' lattice dimensionality solving the system of
Eqs.~(11) with $L=2,3,$ and 4 and treating $d$ in it as an
unknown continuous variable $d^*$.
(For the fully anisotropic $2\times2\times\infty$ Ising lattice,
there is an exact analytical solution~\cite{15}.)
The conclusion is $d^*$ does not depend on $J_x/J_z$
and its plot on $J_y/J_x$ is also presented in Fig.~1.
This plot has a more qualitative character because in the
calculation a cluster with an extremely small size $L=2$ has
been used.
Nevertheless, the presented dependence indicates that the
lattice dimensionality $d^*$ is less than three by
$J_y/J_x<0.3$.

As mentioned in Sec.~III, the energy-energy inverse correlation
length $\kappa_{2,L}$ (and hence the amplitude $A_e$) has a
false behavior in the limit $J_y/J_x\rightarrow0$ due to finite
sizes $L$.
The scaling amplitude $A_f$ obtained from Eqs.~(11) with $L=3$
and 4 suffers from a similar defect.
By finding of $A_e$ and $A_f$, it is not allowed to change
the order of the limits $L\rightarrow\infty$ and
$J_y/J_x\rightarrow0$.
(Note in passing that the calculation of $\kappa_{1,L}$ and
$A_s$ is free upon such requirement.)
Taking into account these circumstances, let us consider in
Fig.~2 the obtained dependencies of ratios $A_e/A_s$ and
$A_eA_f/A_s^2$.
The plots of both dependencies have the horizontal sections
by small deviations of $J_y/J_x$ from unity.
Thus, the amplitude ratios do not depend on the second
anisotropy parameter $J_y/J_x$ in this region of its values.
As $J_y/J_x$ is decreased, both quantities tend to the incorrect
limits.

The recognized properties of the critical FSS amplitudes
by a given orientation ($\alpha$) of an
$L\times L\times\infty$ bar in the anisotropic Ising lattice
can be described by equations
\begin{equation}
   \kappa_{i,L,\alpha}(0, 0) = L^{-1}R_\alpha X_i(0,0)
\end{equation}
and
\begin{equation}
   f_{L,\alpha}^{(s)}(0, 0) = L^{-d}R_\alpha Y(0,0)\;,
\end{equation}
where $X_i(0,0)$ and $Y(0,0)$ are amplitudes of the isotropic
model and $R_\alpha=R_{\alpha}(J_x/J_z, J_y/J_x)$.
The given equations are true at $J_y/J_x=0$ and, according to
the presented data, when $J_y/J_x\rightarrow1$.
Equations (20) and (21) are likely to be valid also over
the wider range of $J_y/J_x$.
This is confirmed qualitatively by the calculation of $A_f$
from Eqs.~(11) with $L=2,3,$ and 4 (without supposition that
$d=3$ for all $J_y\neq0$).

Discuss now the behavior of the directional geometric mean
of the spin-spin inverse correlation length amplitude (table II).
In the two-dimensional case (column with $J_y/J_x = 0$),
$\bar A_s$ loses a stability when $J_x/J_z \leq 10^{-2}$.
This is obviously connected with small widths of strips by which
we approximate the system.
Situation is perceptibly better in three dimensions.
Here $\bar A_s = 1.7(3)$, i.\ e.\ the percentage error equals 18\%.
With such accuracy, we may consider $\bar A_s$ as a constant.


\section{Conclusions}
In this paper, the TM-FSS calculations of critical temperatures,
exponents, amplitudes, and free-energy background for the fully
anisotropic three-dimensional Ising model have been carried out.
The data obtained allow to make the following inference concerning
the structure of critical FSS amplitudes of the inverse correlation
lengths and the free energy:
Similarly to the two-dimensional case, all lattice-anisotropy
parameters are absorbed in a separate prefactor which is
common for named amplitudes and the directional geometric mean of
which is the unity.


\acknowledgements
The author would like to thank V.~Privman, J.~L.~Cardy, M.~Henkel,
and H.~L.~Richards for helpful correspondence via e-mail.
I am grateful to A.~M.~Sterlin, A.~P.~Protogenov, and M.~G.~Teitelman
for useful discussions.
Special thanks go to A.~M.~Babichenko and A.~N.~Karashtin for
valuable consultations in programming.
Finally, it is a pleasure to acknowledge Ms.~Thesba Johnston
for checking of grammar.


\appendix

\section{Quasidiagonalization of the transfer matrices}
The group ${\rm Z_2\times T\land C_{2v}}$
has an order $g=8L^2$.
Its generating elements are a spin inversion $I$, translations
on one step $t_x$ and $t_y$, and reflections in the symmetry
planes $\sigma_v$ and $\sigma_v^\prime$.
In the transfer matrix space
$|S_{11},S_{12},\ldots,S_{LL}\rangle$, they are defined as
\begin{equation}
   I|S_{11},S_{12},\ldots,S_{LL}\rangle=
   |-S_{11},-S_{12},\ldots,-S_{LL}\rangle\;,
\end{equation}
\begin{eqnarray}
   t_x|S_{11},S_{12},& &\ldots,S_{1L};
   S_{21},S_{22},\ldots,S_{2L};\ldots;
   S_{L1},S_{L2},\ldots,S_{LL}\rangle\nonumber\\ & &
   =|S_{1L},S_{11},\ldots,S_{1L-1};
   S_{2L},S_{21},\ldots,S_{2L-1};
   S_{LL},S_{L1},\ldots,S_{LL-1}\rangle
\end{eqnarray}
\begin{eqnarray}
   t_y|S_{11},S_{12},& &\ldots,S_{1L};
   S_{21},S_{22},\ldots,S_{2L};\ldots;
   S_{L1},S_{L2},\ldots,S_{LL}\rangle\nonumber\\ & &
   =|S_{21},S_{22},\ldots,S_{2L};\ldots;
   S_{L1},S_{L2},\ldots,S_{LL};
   S_{11},S_{12},\ldots,S_{1L}\rangle
\end{eqnarray}
\begin{eqnarray}
   \sigma_v|S_{11},S_{12},& &\ldots,S_{1L};
   S_{21},S_{22},\ldots,S_{2L};\ldots;
   S_{L1},S_{L2},\ldots,S_{LL}\rangle\nonumber\\ & &
   =|S_{L1},S_{L2},\ldots,S_{LL};\ldots;
   S_{21},S_{22},\ldots,S_{2L};
   S_{11},S_{12},\ldots,S_{1L}\rangle
\end{eqnarray}
\begin{eqnarray}
   \sigma_v^\prime|S_{11},S_{12},& &\ldots,S_{1L};
   S_{21},S_{22},\ldots,S_{2L};\ldots;
   S_{L1},S_{L2},\ldots,S_{LL}\rangle\nonumber\\ & &
   =|S_{1L},\ldots,S_{12},S_{11};
   S_{2L},\ldots,S_{22},S_{21};
   S_{LL},\ldots,S_{L2},S_{L1}\rangle\;.
\end{eqnarray}
Other transformations of the group are the corresponding
combinations of above operations.
Multiplying from the left the equations like (A1)--(A5) on
conjugate vectors and taking into account the orthonormality
condition
\begin{equation}
   \langle S_{11},S_{12},\ldots,S_{LL}
   |S_{11}^\prime,S_{12}^\prime,\ldots,S_{LL}^\prime\rangle=
   \delta_{S_{11}S_{11}^\prime}\delta_{S_{12}S_{12}^\prime}\ldots
   \delta_{S_{LL}S_{LL}^\prime}\;
\end{equation}
($\delta_{SS^\prime}=\case{1}{2}|S+S^\prime|$ is a Kronecker symbol),
we find the original representation $\Gamma$ of the group.

All matrices of representation built commute with $V$.
For instance, using Eqs.~(5) and (A1), we have
\begin{eqnarray}
   \langle S_{11},S_{12},\ldots,& & S_{LL}|
   I^{-1}VI
   |S_{11}^\prime,S_{12}^\prime,\ldots,S_{LL}^\prime\rangle
   \nonumber\\ & &
   =\langle -S_{11},-S_{12},\ldots,-S_{LL}|V
   |-S_{11}^\prime,-S_{12}^\prime,\ldots,-S_{LL}^\prime\rangle
   \nonumber\\ & &
   =\langle S_{11},S_{12},\ldots,S_{LL}|V
   |S_{11}^\prime,S_{12}^\prime,\ldots,S_{LL}^\prime\rangle
\end{eqnarray}
so that $[V,I]=0$.
The same is valid for all other transformations of the group.

The traces of matrices built are characters of representation
$\Gamma$.
For the $3\times 3\times\infty$ case, the characters of original
representation together with characters of irreducible
representations $\Gamma^{(1)}$ and $\Gamma^{(2)}$ to which
correspond the subblocks containing the largest eigenvalues are
given in table III.
Using this table and utilizing the formula for counting the
multiplicities with which a given irreducible representation
enters into an original representation (see, e.\ g., Ref.~\cite{16})
\begin{equation}
   a_\mu=\frac{1}{g}\sum_ig_i\chi_i^{(\mu)^*}\chi_i
\end{equation}
($g_i$ is a number of elements in $i$th class, $\chi_i^{(\mu)}$
is a character of element from $i$th class in $\mu$th
irreducible representation, and $\chi_i$ is a character of
element from $i$th class in an original representation) we
find the composition of representation $\Gamma$:
\begin{equation}
   \Gamma=18(\Gamma^{(1)}+\Gamma^{(2)})+\ldots\;.
\end{equation}
It follows from here that in a basis where the representation
$\Gamma$ is completely reducible the transfer matrix of 512-th
order will take a quasidiagonal form in which both subblocks
corresponding to the one-dimensional irreducible representations
$\Gamma^{(1)}$ and $\Gamma^{(2)}$ will have the sizes 18 by 18.

The basis vectors of irreducible representations on which the
transfer matrix takes the discussed block-diagonal form are built
with a help of projection operators~\cite{16}.
In the case of an $L=3$ subsystem, the basis vectors for the
irreducible representations $\Gamma^{(1,2)}$ are
\begin{equation}
\begin{array}{ll}
\varphi_1^{(1,2)}=(u_1\pm u_{512})/\sqrt2
\qquad&
\varphi_2^{(1,2)}=\mathop{{\sum}'}_iG_i(u_8\pm u_{505})/\sqrt6
\\
\varphi_3^{(1,2)}=\mathop{{\sum}'}_iG_i(u_{74}\pm u_{439})/\sqrt6
\qquad&
\varphi_4^{(1,2)}=\mathop{{\sum}'}_iG_i(u_{85}\pm u_{428})/2\sqrt3
\\
\varphi_5^{(1,2)}=\mathop{{\sum}'}_iG_i(u_{2}\pm u_{511})/3\sqrt2
\qquad&
\varphi_6^{(1,2)}=\mathop{{\sum}'}_iG_i(u_{4}\pm u_{509})/3\sqrt2
\\
\varphi_7^{(1,2)}=\mathop{{\sum}'}_iG_i(u_{10}\pm u_{503})/3\sqrt2
\qquad&
\varphi_8^{(1,2)}=\mathop{{\sum}'}_iG_i(u_{28}\pm u_{485})/3\sqrt2
\\
\varphi_9^{(1,2)}=\mathop{{\sum}'}_iG_i(u_{79}\pm u_{434})/3\sqrt2
\qquad&
\varphi_{10}^{(1,2)}=\mathop{{\sum}'}_iG_i(u_{11}\pm u_{502})/6
\\
\varphi_{11}^{(1,2)}=\mathop{{\sum}'}_iG_i(u_{15}\pm u_{498})/6
\qquad&
\varphi_{12}^{(1,2)}=\mathop{{\sum}'}_iG_i(u_{75}\pm u_{438})/6
\\
\varphi_{13}^{(1,2)}=\mathop{{\sum}'}_iG_i(u_{16}\pm u_{497})/6
\qquad&
\varphi_{14}^{(1,2)}=\mathop{{\sum}'}_iG_i(u_{76}\pm u_{437})/6
\\
\varphi_{15}^{(1,2)}=\mathop{{\sum}'}_iG_i(u_{30}\pm u_{483})/6
\qquad&
\varphi_{16}^{(1,2)}=\mathop{{\sum}'}_iG_i(u_{84}\pm u_{429})/6
\\
\varphi_{17}^{(1,2)}=\mathop{{\sum}'}_iG_i(u_{12}\pm u_{501})/6\sqrt2
\qquad&
\varphi_{18}^{(1,2)}=\mathop{{\sum}'}_iG_i(u_{86}\pm u_{427})/6\sqrt2\;,
\end{array}
\end{equation}
where
\begin{equation}
   \begin{array}{l}
   u_1=|1,1,1;1,1,1;1,1,1\rangle,\
   u_2=|1,1,1;1,1,1;1,1,-1\rangle,\
   \ldots\ , \\
   u_{512}=|-1,-1,-1;-1,-1,-1;-1,-1,-1\rangle\;.
   \end{array}
\end{equation}
The plus and minus signs correspond to the basis vectors of
irreducible representations $\Gamma^{(1)}$ and $\Gamma^{(2)}$,
respectively.
For shortening of a listing, only the $I$-conjugated pairs of
generating orths are shown in Eqs.~(A10).
The numbers of orths in a pair ($n$ and $n^\prime$) are connected
by a relation $n^\prime=2^N+1-n$.
Acting on such orths by operators
$G_i\in{\rm T\land C_{2v}}$ and taking on each step
only the new $u$-orths (this peculiarity is marked by prime on
the sum symbol), we obtain the expressions for the basis functions
in explicit form.

Finally, having the basis vectors for the irreducible
representations, one can find the matrix elements of subblocks
with the transfer matrix eigenvalues under search.
For the $3\times 3\times\infty$ task, the matrix elements of
subblocks corresponding to the irreducible representations
$\Gamma^{(1,2)}$ have been given with all necessary coefficients
in Ref.~\cite{15}.

In the case of $4\times 4\times\infty$ subsystem, the basis
vectors of $\Gamma^{(1)}$ and $\Gamma^{(2)}$ can be taken in
the form
\begin{equation}
   \begin{array}{ll}
   \psi_1^{(1,2)}=(e_1\pm e_{65\,536})/\sqrt2
   \qquad&
   \psi_2^{(1,2)}=\mathop{{\sum}'}_iG_i(e_2\pm e_{65\,535})/4\sqrt2
   \\
   \ldots \\
   \psi_{671}^{(1,2)}=\mathop{{\sum}'}_iG_i(e_{13\,670}
   \pm e_{51\,867})/4\sqrt2
   \qquad&
   \psi_{672}^{(1,2)}=\mathop{{\sum}'}_iG_i(e_{13\,674}
   \pm e_{51\,863})/8\sqrt2\\
   \\
   \psi_{673}^{(1)}=\case{1}{2}\mathop{{\sum}'}_iG_ie_{256}
   \qquad\qquad\ldots &
   \psi_{787}^{(1)}=(e_{23\,131}+e_{42\,406})/\sqrt2\;,
   \end{array}
\end{equation}
where
\begin{equation}
   e_1=|1,1,\ \ldots\ ,1\rangle,\qquad e_2=|1,1,\ \ldots\ ,-1\rangle,
   \ \ldots\ ,\
   e_{65\,536}=|-1-,1,\ \dots\ ,-1\rangle\;.
\end{equation}
The basis functions (A12) from 1 to 672 and then from 673
to 787 are ordered with the numbers of the first generating
$e$-orths increasing.
Using Eqs.~(5), (A12) and (A13), we evaluate
the matrix elements
$V_{ij}^{(1,2)}=\psi_i^{(1,2)^+}V\psi_j^{(1,2)}$ for subblocks
corresponding to the irreducible representations $\Gamma^{(1,2)}$.
The matrix elements are
\begin{equation}
   V^{(1)}_{ij}=\frac{\max(n_i,n_j)}{\sqrt{n_in_j}}
   \left[g_0^{(i,j)}+2\sum^8_{s=1}g^{(i,j)}_s
   \cosh(2sK_z)\right]
   \exp[\case{1}{2}(m^a_i+m^a_j)K_x
   +\case{1}{2}(m^b_i+m^b_j)K_y]
\end{equation}
and
\begin{equation}
   V^{(2)}_{ij}=2\frac{\max(n_i,n_j)}{\sqrt{n_in_j}}
   \left[\sum^8_{s=1}\tilde g^{(i,j)}_s
   \sinh(2sK_z)\right]
   \exp[\case{1}{2}(m^a_i+m^a_j)K_x
   +\case{1}{2}(m^b_i+m^b_j)K_y]\;,
\end{equation}
where $n_i$ are lengths of basis vectors, $m_i^a$ and $m_i^b$
are the reduced partial energies of spin configurations in orths
of $i$th vector.
All coefficients $g_s^{(i,j)}$ are non-negative
and satisfy to the ``sum rules''
\begin{equation}
   g_0^{(i,j)}+2\sum_{s=1}^8g_s^{(i,j)}=\min(n_i,n_j)\;.
\end{equation}
We did not keep the coefficients $g_0^{(i,j)}$ but
restored them for each matrix element $V_{ij}^{(1)}$ from
Eqs.~(A16).
As a calculation shows, the coefficients $g_s^{(i,j)}$ with
$s\neq 0$ are not greater than 60.
Hence, it is enough to take one byte for every element of the
$g$-array, i.\ e.\ to use the data type `char' in {\it C\/} code.
Thus, it is required $2\,480\,624$ bytes of a memory to
store the $g$-coefficients for a triangle part of symmetric matrix
$V^{(1)}$.
The values of coefficients $\tilde g_s^{(i,j)}$ lie in the range
from $-28$ to $+40$ and we alloted in addition the $1\,809\,024$
bytes of a memory for the $\tilde g$-coefficients of matrix
$V^{(2)}$.


\section{Formulae for the calculation of susceptibilities}
In deriving of formulae for $\chi_L$, we will point out from
a fluctuation-dissipation relation connecting the susceptibility
with a magnetic moment ${\cal M}$ (see,for example, Ref.~\cite{17}):
\begin{equation}
   \chi_L(T)=\frac{1}{k_BT}\lim_{M\to\infty}
   \frac{1}{L^2M}
   \langle{\cal M}^2\rangle\;.
\end{equation}
Here ${\cal M}=\sum\nolimits_{ijk}S_{ij}^k$ where
$S_{ij}^k\equiv S_{ijk}$ is the total magnetic
moment of $L\times L\times M$ periodic subsystem; the brackets refer
to average on Gibbs distribution.
Taking into account the translational invariance of a cluster
in the longitudinal ($z$) direction, one can write Eq.~(B1) in
the form
\begin{equation}
   \chi_L(T)=\frac{1}{L^2k_BT}\lim_{M\to\infty}
   \sum_{r=0}^{M-1}\langle(S_{11}^k+S_{12}^k+\ldots
   +S_{LL}^k)(S_{11}^{k+r}
   +S_{12}^{k+r}+
   \ldots+S_{LL}^{k+r})\rangle\;.
\end{equation}
To calculate the statistical means, we use the transfer matrix
technique.
Let us introduce in addition the spin matrices making by this
the one-dimensional order of pair of indexes
$i,j\rightarrow l=L(i-1)+j$:
\begin{equation}
   \hat S_l=\underbrace{\mathstrut 1\times \ldots\times1}_{l-1}
   \times\sigma_z\times
   \underbrace{1\times\ldots\times1}_{N-l}\;,
\end{equation}
where $1$ denotes the unit matrix of second order and $\sigma_z$
is Pauli's $z$-matrix; $N=L^2$.
This allows to rewrite Eq.~(B2) as
\begin{equation}
   \chi_L(T)=\frac{1}{L^2k_BT}\lim_{M\to\infty}
   \frac{1}{{\rm Tr}V^M}
   \sum_{r=0}^{M-1}{\rm Tr}[(\hat S_1+\ldots
   +\hat S_N)V^r(\hat S_1+\ldots
   +\hat S_N)V^{M-r}]\;.
\end{equation}
 From here, by passing under trace symbol into diagonal
representation of the transfer matrix and by taking
into account the non-degeneracy of its largest eigenvalue,
we obtain
\begin{equation}
   \chi_L(T)=\frac{1}{L^2k_BT}\sum_{i=1}^{2^N-1}
   \frac{\Lambda_0+\Lambda_i}{\Lambda_0-\Lambda_i}
   |F_i^+(\hat S_1+\ldots+\hat S_N)F_0|^2\;,
\end{equation}
where $F_0,F_1,\ldots$ are eigenvectors of matrix $V$
corresponding to its eigenvalues $\Lambda_0,\Lambda_1,\ldots$.
Further, the operator $\hat S=\hat S_1+\ldots+\hat S_N$
is invariant with respect to all purely spatial transformations
and breaks the ${\rm Z}_2$ symmetry.
Therefore, the matrix elements entering into Eq.~(B5)
are not zero only for ``transitions'' from the identity
irreducible representation $\Gamma^{(1)}$ just into
the irreducible representation $\Gamma^{(2)}$.

Vector $F_0$ is a linear combination of basis functions
only of the identity irreducible representation.
In the case of $3\times 3\times\infty$,
\begin{equation}
   F_0=\sum_{i=1}^{18}f_i^{(0)}\varphi_i^{(1)}\;,
\end{equation}
where $f_i^{(0)}$ are components of eigenvector answering to the
largest eigenvalue ($\Lambda_0$) of subblock of the identity
irreducible representation.
Using Eqs.~(A10), we find that
\begin{equation}
   \hat S\varphi_i^{(1)}=m_i\varphi_i^{(2)}\;,
\end{equation}
where
\begin{equation}
   m_i=\{9,3,3,3,7,5,5,1,1,5,3,3,1,1,1,1,3,1\}\; ;
\end{equation}
$m_i$ is the magnetic moment of spin configurations in the $i$th
basis vector.
As a result, we obtain from Eq.~(B5) the following work formula
for a calculation of the susceptibility:
\begin{equation}
   \chi_3(T)=\frac{1}{9k_BT}\sum_{i=1}^{18}
   \frac{\Lambda_0+\Lambda_i^{(2)}}{\Lambda_0-\Lambda_i^{(2)}}
   \left[\sum_{j=1}^{18}m_jf_j^{(0)}f_j^{(i)}\right]^2\;.
\end{equation}
Here $f_j^{(i)}$ are components of $i$th eigenvector
corresponding to eigenvalue $\Lambda_i^{(2)}$ for the subblock of
irreducible representation $\Gamma^{(2)}$.

Analogous formula take place for the $L=4$ subsystem:
\begin{equation}
   \chi_4(T)=\frac{1}{16k_BT}\sum_{i=1}^{672}
   \frac{\Lambda_0+\Lambda_i^{(2)}}{\Lambda_0-\Lambda_i^{(2)}}
   \left[\sum_{j=1}^{672}m_jf_j^{(0)}f_j^{(i)}\right]^2\;.
\end{equation}
All quantities entering in this expression should be taken,
of course, for the $4\times 4\times\infty$ model.

Therefore, the calculation of susceptibilities requires the solution
of a part eigenproblem for the subblock of an identity irreducible
representation and the solution of a full eigenproblem
for a second subblock which corresponds to the irreducible
representation $\Gamma^{(2)}$.
The part eigenproblem was solved again by the conjugate gradient
method and the full one --- by using the library {\it C\/} pair
{\it tred2 - tqli\/}.



\begin{table}
\caption{Critical temperatures, background, critical FSS amplitudes
and their ratios for different values of the anisotropy
parameters $J_x/J_z$ and $J_y/J_x$.}
\begin{tabular}{llccccccc}
$J_y/J_x$&$J_x/J_z$&$k_BT_c/J_z$&$A_s$&$A_e$&$A_e/A_s$&$A_f$&$f_0$&
$A_f/A_s$\\
\tableline
 1.0 &1.0  &4.58104&1.4401&4.9627&3.44&0.4189&0.773&0.290\\
     &0.1  &1.35037&0.3613&1.2847&3.55&0.1044&0.959&0.288\\
     &0.01 &0.65458&0.0723&0.2576&3.56&0.0208&1.576&0.287\\
     &0.001&0.40917&0.0115&0.0411&3.57&0.0033&2.451&0.286\\ \\
 0.75&1.0  &4.18009&1.3345&4.5795&3.43&0.3953&0.775&0.296\\
     &0.1  &1.27931&0.3317&1.1712&3.53&0.0973&0.985&0.293\\
     &0.01 &0.63312&0.0653&0.2310&3.53&0.0191&1.623&0.292\\
     &0.001&0.39985&0.0103&0.0365&3.54&0.0030&2.508&0.291\\ \\
 0.5 &1.0  &3.73973&1.2288&4.0812&3.32&0.3924&0.782&0.319\\
     &0.1  &1.19903&0.3005&1.0251&3.41&0.0943&1.019&0.313\\
     &0.01 &0.60815&0.0580&0.1981&3.41&0.0181&1.683&0.312\\
     &0.001&0.38882&0.0090&0.0309&3.43&0.0028&2.578&0.311\\ \\
 0.25&1.0  &3.22427&1.1256&3.3542&2.97&0.4407&0.803&0.391\\
     &0.1  &1.10117&0.2665&0.8151&3.05&0.1016&1.073&0.381\\
     &0.01 &0.57655&0.0500&0.1533&3.06&0.0190&1.767&0.380\\
     &0.001&0.37453&0.0076&0.0235&3.09&0.0029&2.675&0.381\\ \\
 0.0 &1.0  &2.32081&0.8917&5.9901&6.71&0.2952&0.914&0.331\\
     &0.1  &0.91079&0.1856&1.3661&7.36&0.0616&1.232&0.331\\
     &0.01 &0.51058&0.0327&0.2418&7.39&0.0108&1.983&0.330\\
     &0.001&0.34346&0.0048&0.0349&7.27&0.0016&2.915&0.333\\
\end{tabular}
\end{table}


\newpage
\begin{table}
\caption{Directional geometric mean of the spin-spin inverse
 correlation length amplitude ${\bar A}_s$ by different values of
 $J_x/J_z$ and $J_y/J_x$.}
\begin{tabular}{lccccc}
&&&$J_y/J_x$&&\\
$J_x/J_z$\\ \cline{2-6}
&0&0.25&0.5&0.75&1.0\\
\tableline
1.0   &0.891 &1.57 &1.46 &1.43 &1.44\\
0.1   &0.833 &2.00 &1.87 &1.80 &1.76\\
0.01  &0.577 &2.02 &2.01 &2.01 &2.02\\
0.001 &0.294 &1.48 &1.51 &1.54 &1.57\\
\end{tabular}
\end{table}


\newpage
\begin{table}
\caption{Characters of the group
 ${\rm Z_2\times T\land C_{2v}}$ in the case $L=3$;
 here ${\rm T\land C_{2v}\approx
 C_{3v}\times C_{3v}}$ .}

\begin{tabular}{lllllll}
&$E$&$3\sigma_v$
&$9\sigma_v\sigma_v^\prime$
&$2t_x$&$6t_x\sigma_v$&
$I, 3I\sigma_v, 3I\sigma_v^\prime$\\
&&$3\sigma_v^\prime$
&&$2t_y$&$6t_y\sigma_v^\prime$&
$9I\sigma_v\sigma_v^\prime, 2It_x, 2It_y$\\
&&&&$4t_xt_y$&&
$4It_xt_y, 6It_x\sigma_v,
6It_y\sigma_v^\prime$\\
\tableline
$\Gamma^{(1)}$&1&1&1&1&1&1\\
$\Gamma^{(2)}$&1&1&1&1&1&-1\\
\\
$\Gamma$&512&64&32&8&4&0\\
\end{tabular}
\end{table}


\begin{figure}
 \caption{Critical exponents $\nu$ and $\gamma/\nu$ (left scale)
 and the effective lattice dimensionality $d^*$ (right scale)
 versus anisotropy parameter $J_y/J_x$.}
\end{figure}
\begin{figure}
 \caption{The amplitude ratios $A_e/A_s$ and $A_eA_f/A_s^2$
 against the anisotropy parameter $J_y/J_x$. The curve parts
 which are considered as non-physical ones are shown by dashed
 line.}
\end{figure}

\end{document}